\documentclass[twocolumn,aps,nofootinbib]{revtex4-1}
\usepackage{amssymb}
\usepackage{amsmath}
\usepackage{epsfig}
\usepackage{tabularx}
\usepackage{inputenc}
\usepackage{url}
\usepackage{breqn}
\usepackage{hyperref}
\usepackage{footmisc}
\usepackage{titlesec}
\usepackage{fancyhdr}
\usepackage{graphicx}
\usepackage{caption}
\usepackage{slashed}
\usepackage{wrapfig}
\usepackage{color}
\usepackage{mathrsfs}
\graphicspath{{latex_images/}}
\begin{document}
\begin{titlepage}
\begin{center}
\hfill  MI-TH-1881
\end{center}
\title{Probing a simplified, $W^{\prime}$ model of $R(D^{(\ast)})$ anomalies using $b$-tags, $\tau$ leptons and missing energy}
\vspace{1.0cm}
\author{\textbf{Mohammad Abdullah$^{\bf a}$, Julian Calle$^{\bf c}$, Bhaskar Dutta$^{\bf a}$, Andr\'es Fl\'orez$^{\bf b}$, Diego Restrepo$^{\bf c}$} \\
\vspace{1.0cm}
\normalsize\emph{$^{\bf a}$Mitchell Institute for Fundamental Physics and Astronomy, Department of Physics  and Astronomy, Texas A$\&$M University,
College Station, TX 77843}\\
\normalsize\emph{$^{\bf b}$Physics Department, Universidad de los Andes, Bogot\'a, Colombia}\\
\normalsize\emph{$^{\bf c}$Instituto de F\'\i sica, Universidad de Antioquia,
  Calle 70 No. 52-21, Apartado Aéreo 1226, Medellín, Colombia}\\
\vspace{1.5cm}
}
\begin{abstract}
We study the LHC sensitivity to a $W'$ produced via bottom and charm quarks and decaying to $\tau$ flavor leptons in the mass range 200-1000 GeV. We show that the extra $b$ quarks necessitated by the production mechanism can improve the background rejection compared to an inclusive analysis relying solely on $\tau$-tagging and $E_{T}^{\text{miss}}$. We present prospective limits on the couplings and compare them to the best fit to the $R(D^{(\ast)})$ anomalies in $B$ meson decays.
\end{abstract}

\maketitle
\end{titlepage}

\section{Introduction}
\label{sec:intro}
Recently, the BaBar \cite{Lees:2013uzd,Lees:2012xj}, Belle \cite{Huschle:2015rga,Abdesselam:2016cgx,Sato:2016svk,Hirose:2016wfn}, and LHCb \cite{Aaij:2015yra} collaborations have measured the semi-leptonic decays of $B$-mesons to $D$ and $D^\ast$ and found a sizable discrepancy from the standard model (SM) value. More specifically, the anomaly is manifest in the observables
\begin{eqnarray}
R(D) &=& \frac{\operatorname{Br}(\bar{B}\rightarrow D \tau^- \bar{\nu}_\tau)}{\operatorname{Br}(\bar{B}\rightarrow D \ell^- \bar{\nu}_\ell)}\\
R(D^\ast) &=& \frac{\operatorname{Br}(\bar{B}\rightarrow D^\ast \tau^- \bar{\nu}_\tau)}{\operatorname{Br}(\bar{B}\rightarrow D^\ast \ell^- \bar{\nu}_\ell)},
\end{eqnarray}
where $\ell=e,\; \mu$. The denominator is averaged over electrons and muons in BaBar and Belle, while for LHCb only muons contribute. The special property of these observables is that, in the SM, the hadronic factors are expected to cancel out which reduces the uncertainty in our prediction. 
The measurements are found to disagree with the SM predictions at about 4$\sigma$~\cite{Wormser:2017hsx}. 

The descripency  suggests lepton flavor universality violation and has prompted many works to explain the anomaly in the context of new physics models~\cite{Asadi:2018wea,He:2017bft,Biswas:2017vhc,Akeroyd:2017mhr,Choudhury:2017qyt,Iguro:2017ysu,Watanabe:2017mip,Dutta:2017wpq,Alok:2017qsi,DiLuzio:2017chi,Matsuzaki:2017bpp,
Altmannshofer:2017poe,Choudhury:2017ijp,Faroughy:2016osc,Crivellin:2017zlb,Megias:2017ove,Chen:2017eby,Chen:2017hir,Hiller:2016kry,Ko:2017lzd,
Barbieri:2016las,Cvetic:2017gkt,Wei:2017ago,Popov:2016fzr,Wang:2016ggf,Bhattacharya:2016mcc,Becirevic:2016yqi,
Sahoo:2016pet,Kim:2016yth,Deshpand:2016cpw,Boucenna:2016qad,Li:2016vvp,Das:2016vkr,Boucenna:2016wpr,Zhu:2016xdg,Cline:2015lqp,Becirevic:2016oho,
Barbieri:2015yvd,Fajfer:2015ycq,Hati:2015awg,Bauer:2015knc,Bordone:2017anc,Celis:2016azn,Alonso:2016oyd,Ligeti:2016npd,
Bhattacharya:2016zcw,Bardhan:2016uhr,Ivanov:2016qtw,Nandi:2016wlp,Alonso:2016gym,Bhattacharya:2015ida,Freytsis:2015qca,
Calibbi:2015kma,Greljo:2015mma,Alonso:2015sja,Biancofiore:2013ki,Fajfer:2012jt,Datta:2012qk,Bailey:2012jg,Tanaka:2012nw,Blanke:2018sro,Greljo:2018ogz,Dasgupta:2018nzt}.
The underlying interaction  that accomodates the experimental result arises from the charged current mediated decay $b\rightarrow c\tau\nu$, which is CKM-suppressed in the SM. One simple way to obtain a new physics contribution to this charged current  is to use a $W^\prime$ gauge boson  which couples to  the second and third generation  fermions~\cite{Greljo:2015mma}. In order to explain the anomaly, the $W^\prime$ does not need to couple to the first generation which seemingly makes the model harder to explore at the LHC. In addition, the existence of $\tau$s in the final state  also leads to a difficulty in overcoming the SM background at the LHC . 

In this paper, we investigate a $W^\prime$ which couples only to bottom and charm quarks in the color sector and to $\tau$ flavor in the lepton sector. The dominant production channels in this case are $gg$ and $gc$ fusion, both of which lead to an associated $b$-jets in the final state. The presence of this $b$-jet in the final state opens up a possibility of mitigating the SM background by performing an exclusive analysis that requires at least one  $b$-jet in the final state along with a hadronic $\tau$ and missing energy, potentially improving our efficiency. In addition, the existence of a $b$-jet  would help us deduce the existence of a  third generation coupling of $W^\prime$ which is crucial for the explanation of the anomaly. 
To better understand the effectiveness of this technique, we will contrast our use of $b$-tags with reinterpretations of inclusive $W^\prime$ analyses done by ATLAS and CMS in which no such $b$-jet requirements are made. 

In Refs. \cite{Khachatryan:2015pua,Aaboud:2018vgh}, the inclusive search of $W^\prime$ is done without reference of the production mechanism and therefore without the $b$-jet mandate. Ref.~\cite{Altmannshofer:2017poe} considers the $gc\rightarrow b\tau$ process at the LHC but only as an effective vertex and by looking for leptonic $\tau$ decays rather than hadronic. Their results are not applicable to the mass range we are interested in. In Ref.~\cite{Asadi:2018wea,Greljo:2018ogz} limits on $W^\prime$ are set in the context of a complete model and are therefore highly model dependent. In Ref. \cite{Dalchenko:2017shg}, the importance of $b$-jets in resonance searches is demonstrated but using a $Z^\prime$ gauge boson decaying to muons.    

We should point out that, regardless of $B$ anomalies, the properties mentioned above are worthy of investigation in their own rights, in addition to serving as an additional jigsaw puzzle piece in the LHC resonance search program \cite{Craig:2016rqv}. 

The rest of this paper is organized as follows: in Section \ref{sec:model} we present our model and refer to related works in the literature. In \ref{sec:RD} we give a brief presentation of the $R(D^{(\ast)})$ anomalies and how they can be accommodated in our model. In \ref{sec:samples} we describe our setup, in \ref{sec:analysis} we explain our search methodology, in \ref{sec:results} we present our results, and finally, in \ref{sec:conclusion} we conclude.

\section{Model and collider phenomenoligy}
\label{sec:model}

The interactions connecting the new physics in our model to the SM are described by the following Lagrangian:
\begin{equation}
\mathcal{L} = (g_q' \bar{c}\gamma^\mu P_L b + g_\tau ' \bar{\nu}_\tau \gamma^\mu P_L \tau^- ){W'}^{+}_\mu+h.c.
\label{eq:lag}
\end{equation}
where $g_q'$ and $g_\tau'$ are new physics couplings. Due to the flavor violating nature of these couplings, constructing a UV completion that is consistent with the SM is not straight forward. An examples where such a UV completion has been discussed is Ref. \cite{Boucenna:2016qad}, where the $W'$ is assumed to only couple directly to a generation of vector-like fermions too heavy to observe. The couplings in Eq. \eqref{eq:lag} are then induced via mixings between the new fermions and SM fermions. In any such UV model, plenty of extra assumptions are made that need to be individually tested against experiment, and one can rarely rule out the entire space of possible models. Therefore, rather than demonstrate the viability of this particular instance of such models, we adopt the pragmatic approach of referring to the aforementioned paper as a proof of principle and focus instead on the collider implications on a minimal construct.

\begin{figure}
\begin{center}
\includegraphics[width = 0.27 \textwidth]{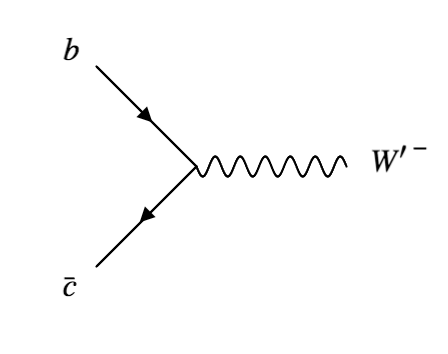}
\includegraphics[width = 0.34 \textwidth]{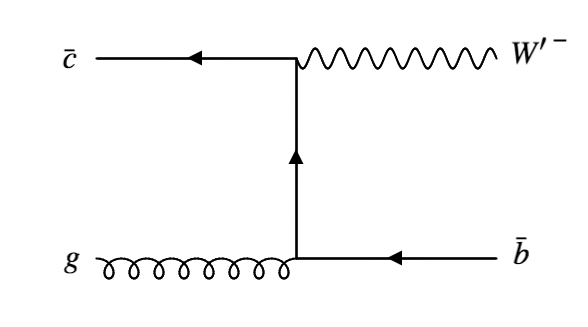}
\includegraphics[width = 0.34 \textwidth]{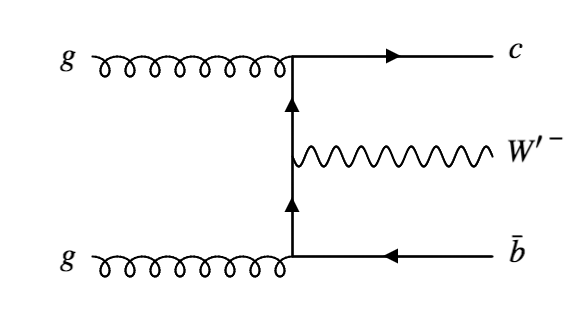}
\caption{Representative Feynman diagrams of $W'$ production at the LHC.}
\label{fig:diag}
\end{center}
\end{figure}

The main production channels at the LHC are shown in Fig. \ref{fig:diag}. Production via any other initial state is suppressed either by extra couplings or by highly off-shell intermediate states. While a larger order of couplings and number of final states would order the cross section contributions of these diagrams from top to bottom, this is counteracted by the difference in momentum fraction carried by gluons, charm quarks, and bottom quarks. One would naively expects the diagrams with either gluons or charms in the initial state to be dominant. As we will see, the efficiency of the $b$-tag requirement suggests that this is true to a good extent. 

Since we are interested in masses of 200 GeV and above, all the decay products are effectively massless leading to very simple expressions for the branching ratios:
\begin{eqnarray}
\operatorname{Br}(W' \rightarrow bc) &\simeq& \frac{3{g'}_q^2}{3{g'}_q^2+{g'}_\tau^2}\\
\operatorname{Br}(W' \rightarrow \tau \nu) &\simeq& \frac{{g'}_\tau^2}{3{g'}_q^2+{g'}_\tau^2}
\end{eqnarray}

Note: The \textsc{FeynRules} model files,  \textsc{Madgraph} process cards, and the \textsc{Mathematica} notebook used for the study are available online~\cite{mo_abdullah_2018_1240452}
for the purposes of reproducibility, utility, and peer review.
\section{$R(D^{(\ast)})$ anomalies}
\label{sec:RD}

As mentioned in the introduction, the BaBar, Belle and LHCb collaborations have measured $R(D)$ and $R(D^\ast)$ to very high precision.
Using the results of Refs. \cite{Bigi:2016mdz,Jaiswal:2017rve} (see also \cite{Fajfer:2012vx}) the theory values are:
\begin{eqnarray}
R(D)_{\text{SM}} &=& 0.298 \pm 0.003,\\
R(D^\ast)_{\text{SM}} &=&  0.255 \pm 0.004,
\end{eqnarray}
while the combined experimental values are \cite{HFAG}:
\begin{eqnarray}
R(D)_{\text{Exp}} &=& 0.407 \pm 0.039 \pm 0.024, \\
R(D^\ast)_{\text{Exp}} &=&  0.304 \pm 0.013 \pm 0.007.
\end{eqnarray}

Following \cite{Boucenna:2016qad}, the observables above are modified as follows:
\begin{equation}
R(D^{(\ast)})_{\text{NP}} = \left(1+\frac{g'_q g'_\tau}{m_{W'}^2}\frac{\sqrt{2}}{4 G_F V_{cb}}\right)^2 R(D^{(\ast)})_{\text{SM}},
\end{equation}
%
where $m_{W'}$ is the $W'$ mass, $G_F=1.16\times 10^{-5}\; \text{GeV}^{-2}$ is the Fermi constant, and $V_{cb} = 0.04$ is the $cb$ component of the CKM matrix. Taking both new couplings to be positive, the central values of $R(D)$ and $R(D^\ast)$ require the factor $g'_q g'_\tau/m_{W'}^2$ to be $0.002\, (100\, \text{GeV}/m_{W'})^2$ and $0.001\,(100\, \text{GeV}/m_{W'})^2$ respectively. In this study we will present our limits in the $m_{W'}-g'_q$ plane for several representative values of $g'_\tau$

\section{Samples and Simulation}
\label{sec:samples}

The production of top quark pairs ($t\bar{t}$) with associated jets from initial state radiation (ISR), commonly referred to as semi-leptonic  $t\bar{t}+\text{jets}$ events, where $t\bar{t} \to bW^{+}\bar{b}W^{-} \to b\bar{b}\ell^{\pm}\nu_{\ell}q\bar{q}$,  represent the dominant source of background (61\% of the total background rate). The next source of background, 21\%, comes from the production of a $W$ boson with associated ISR jets ($W$+jets), where a real prompt lepton from the $W$ is produced and missing transverse energy ($E^{\text{miss}}_{T}$) is obtained from the associated neutrino from the $W$ decay. Production of Drell-Yann (DY) events ($Z/\gamma^{*}$), plus ISR jets, accounts for roughly 11\% of the total background. 
Leptonic decays of the $Z/\gamma^{*}$ bosons enter in the final event selection when one of the leptons is lost in the reconstruction, becoming missing transverse energy. The remaining contributions come from events with single tops and production of vectors boson pairs ($WW, ZZ, WZ$), referred to as diboson events, plus ISR jets. The simulated signal and background samples are LO and were produced using \textsc{MadGraph} (v.2.6.0)~\cite{Alwall:2014hca} interfaced with \textsc{PYTHIA 8} \cite{Sjostrand:2014zea} for parton hadronization and  \textsc{DELPHES} (v3.3.2)~\cite{deFavereau:2013fsa} to include particle interactions with a  detector (CMS configuration card was used). The model implementation was generated using \textsc{FeynRules 2.3} \cite{Christensen:2008py,Degrande:2011ua,Alloul:2013bka} 

For signal events, jet matching and merging was implemented using the MLM algorithm \cite{Alwall:2007fs}. The matching parameters, xqcut and QCUT were checked to result in a reasonably stable cross section and smooth transition in the differential jet rate distribution between events with $N$ and $N+1$ jets, but were not optimized further due to the inherent issues in modeling heavy quark emissions. The xqcut variable defines the minimal distance between partons at the MadGraph level, while the QCUT variable sets the minimum energy spread for a clustered jet in PYTHIA. For the 
signal production an xqcut of 30 and a QCUT of 60 were used with a 5 jet flavor scheme. 

Note that there is a large systematic uncertainty in the jet matching which we do not account for. This uncertainty is larger when the higher multiplicity diagrams (the bottom two diagrams in Fig. \ref{fig:diag}) are dropped even though the matched cross section is unchanged. We do not expect the uncertainty to affect the selection efficiency of the hard jets we require, although the total cross section could in reality be higher or lower by up to 30\%. This does modify our absolute projected sensitivity, but the comparison between the exclusive and inclusive reach remains valid.

Another potential source of error are the collinear divergences at high $W'$ masses in the second the third diagrams which may spoil the validity of the perturbative calculation \footnote{We thank Richard Ruiz for pointing this out and providing valuable advice on how to proceed.}. One way to avoid this is to restrict the $p_T$ cut value of the $b$-jet as prescribed in  \cite{Degrande:2016aje}. Alternatively, we may simulate the events at NLO with proper parton showering to extend the domain of validity of our $p_T$ cut. Following the procedure used by \cite{Fuks:2017vtl} we generate model files compatible with NLO calculations using \textsc{FeynRules 2.3} \cite{Degrande:2014vpa} and produce samples to compare with the LO results. We find that the cross section increases by about 15\% for a 200 GeV $W'$ and about 7\% for 1 TeV at NLO compared to LO, while any alteration to the b-jet p_T distribution is insignificant. We use the LO results to set our sensitivity keeping in mind that the NLO results would enhance our reach.


\section{Analysis}
\label{sec:analysis}

The analysis required at least one hadronic tau, $\tau_{h}$, with $p_{T} > 80$ GeV and $|\eta_{\tau_{h}}| < 2.3$ in order to constrain the selected $\tau_{h}$ candidates to be within the tracking acceptance region. Events with an 
additional $\tau_{h}$ candidate with $p_{T} > 50$ GeV and $|\eta_{\tau_{h}}| < 2.3$ were vetoed, which helped to reduce the diboson and DY+jets contributions. Because of the signal topology, a jet with $p_{T} > 20$ GeV and 
$|\eta| < 2.5$ identified as a $b$-quark is selected. To further suppress backgrounds with different multiplicities of other prompt leptons, events with electrons or muons with $p_{T} > 15$ GeV are not allowed. Since signal events are expected to have 
large $E^{\text{miss}}_{T}$, on average half of the $W'$ mass, a 140 GeV minimum threshold is imposed. Finally, given the expected heavy nature of $W'$ the $\phi$ angle between the $\tau_{h}$ and the $E^{\text{miss}}_{T}$ is expected to be large. Therefore, a $|\Delta \phi (\tau_{h}, E^{\text{miss}}_{T})| > 2.4$ is required, which allows to remove over 85\% of the remaining background while keeping over 75\% of the signal.

\begin{table}
\begin{center}
\caption {Event selection criteria used for analysis.}
\label{tab:selections}
\begin{tabular}{ l  c}\hline\hline
Criterion & Selection\\
 \hline
    $N_{\tau_{h}}$       & $\ge$ 1 \\
    $|\eta_{\tau_{h}}|$ & $< 2.3$ \\
    $p_{T} (\tau_{h})$ & $> 80$ GeV \\
    Veto second $\tau_{h}$ &  $> 50$ GeV \& $|\eta_{\tau_{h}}| < 2.3$ \\
    $N_{e/\mu}$ with $p_{T} > 15$ GeV & $= 0$ \\
    $N_{b-\text{jets}}$ & $= 1$\\
    $p_{T} (b-\text{jets})$ & $> 20$ GeV \\
    $|\eta_{b-\text{jets}}|$ & $< 2.5$ \\
    $ E^{\text{miss}}_{T}$ & $> 140$ GeV \\
     $|\Delta \phi (\tau_{h}, E^{\text{miss}}_{T})|$ & $ > 2.4$ \\
   \hline\hline
 \end{tabular}
\end{center}
\end{table}

\begin{figure}
 \begin{center} 
 \includegraphics[width=0.48\textwidth, height=0.35\textheight]{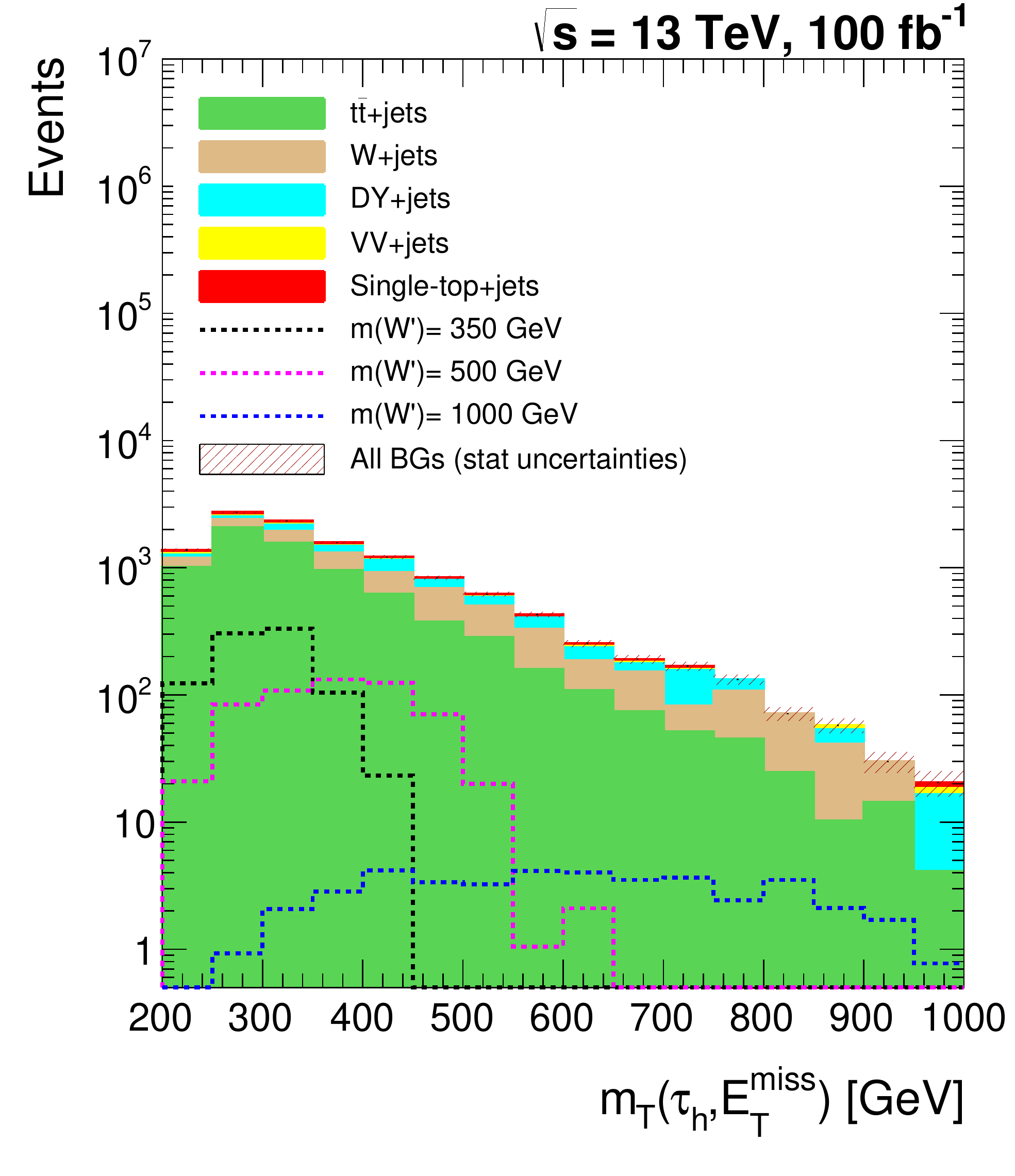}
 \end{center}
 \caption{$m_{T}(\tau_{h},E^{\text{miss}}_{T})$ distribution after all the event selection criteria. The backgrounds are stacked while the signals are overlaid. All the processes included are normalized to the production cross section and luminosity (100 $\text{fb}^{-1}$).}
 \label{fig:mTstacked}
 \end{figure}

By selecting at least one well identified high-$p_{T}$ $\tau_{h}$, a $b$-jet, and $E^{\text{miss}}_{T} > 140$ GeV, background contributions from Quantum Chromodynamic (QCD) processes are deemed negligible. 
Table~\ref{tab:selections} summarises the event selection criteria used for the analysis. Figure~\ref{fig:mTstacked} shows the transverse mass distribution between the $\tau_{h}$ and the $E^{\text{miss}}_{T}$ for the backgrounds and 
three different signal points. The backgrounds and signals are normalised to the corresponding production cross section and 100 $\text{fb}^{-1}$ of luminosity. 


 \section{Results}
\label{sec:results}

In Tables \ref{tab:cutefficiencySMnew} and \ref{tab:cutefficiencynw} we show the  efficiencies of the kinematical selections based on Table \ref{tab:selections} and significances for 100 fb$^{-1}$ of luminosity. We find that  the significance increases as the mass of $W^\prime$ increases from 200 GeV to 400 GeV due to the increase in missing energy in the final state. We also note  that the 250-700 GeV mass range can be probed using 100 fb$^{-1}$ of luminosity at the 5$\sigma$ level for $g'_{q}=g'_\tau=0.1$.

\begin{figure}
\begin{center}
\includegraphics[width = 0.48 \textwidth]{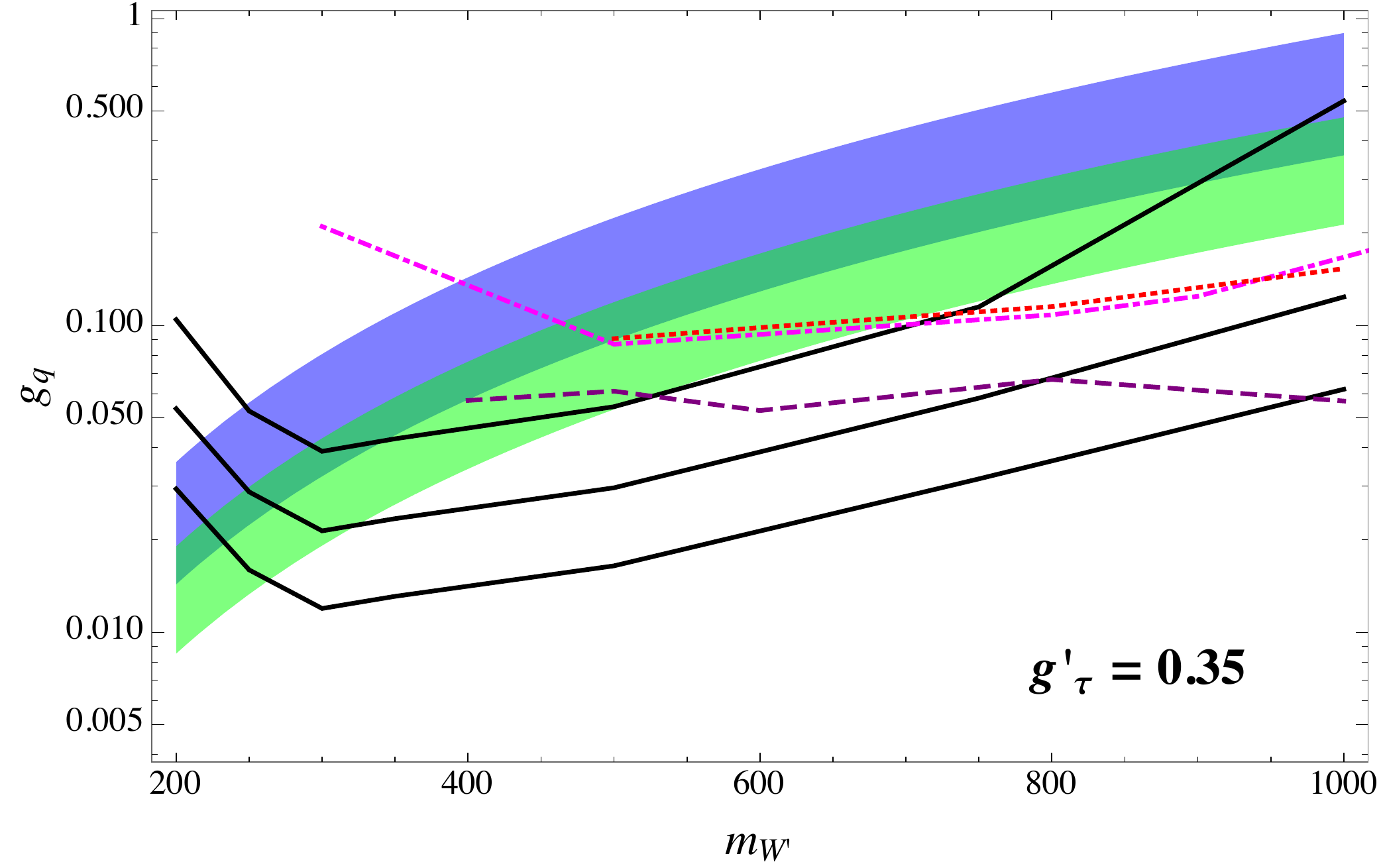}
\includegraphics[width = 0.48 \textwidth]{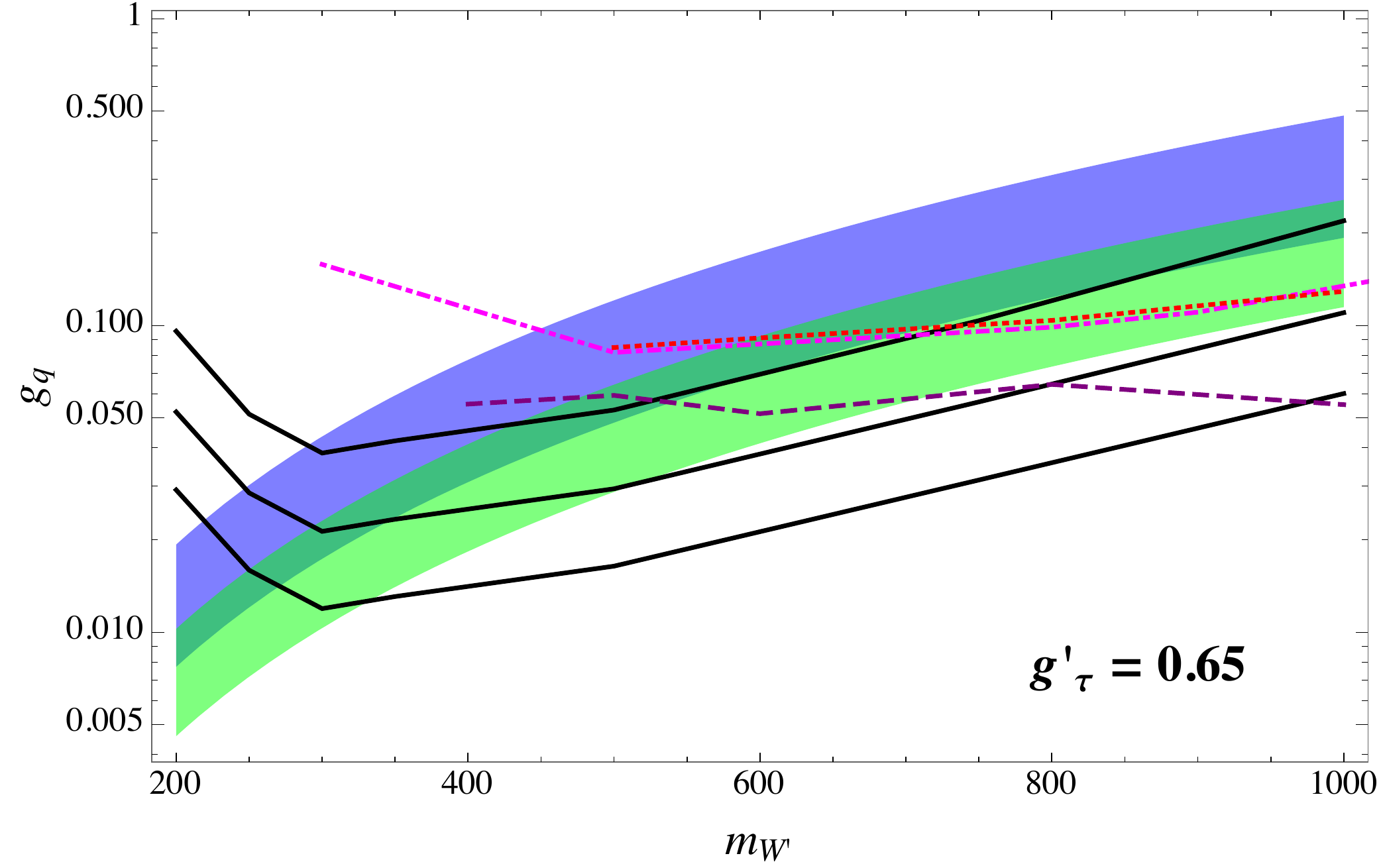}
\includegraphics[width = 0.48 \textwidth]{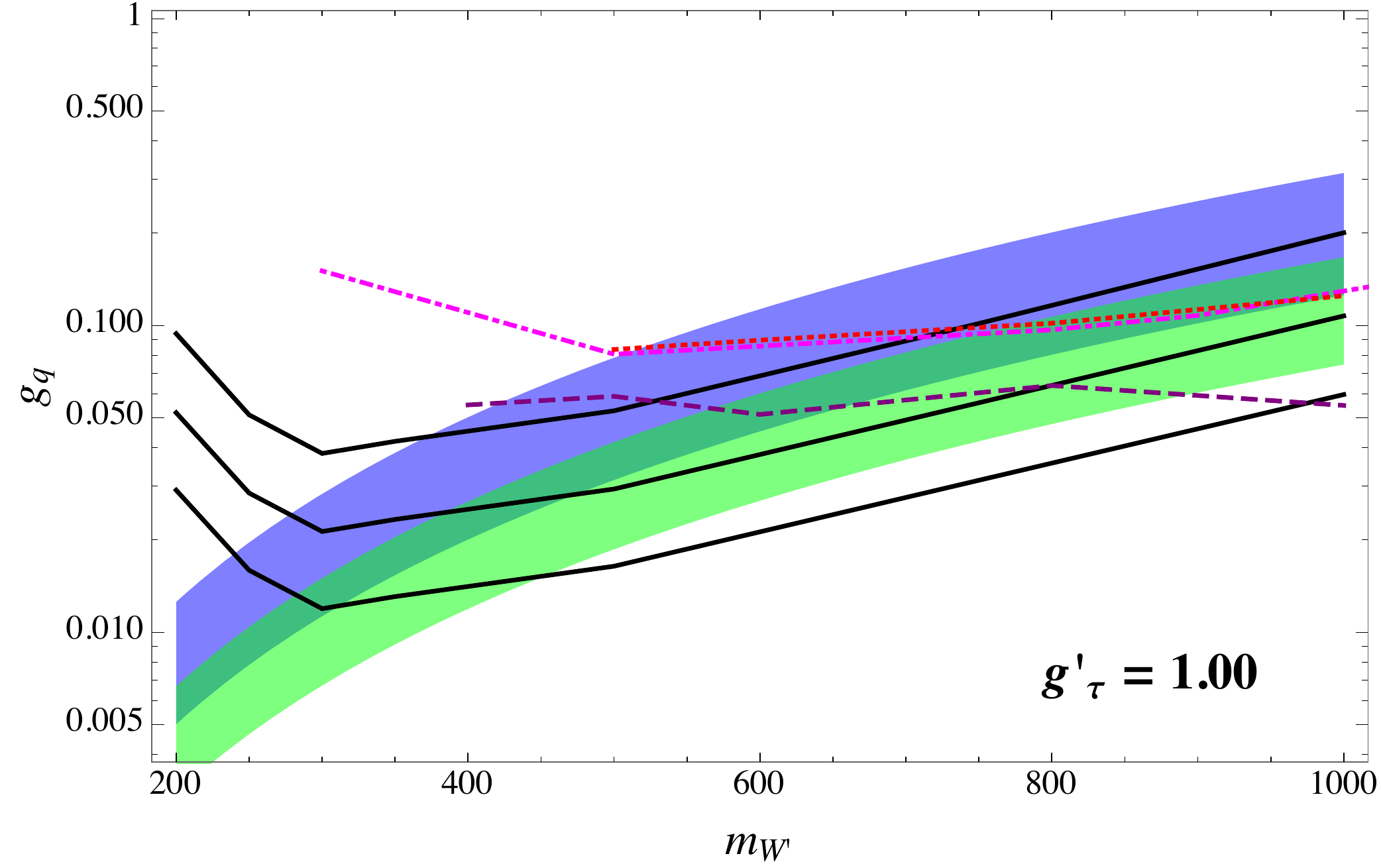}
\caption{Projected 3$\sigma$ sensitivity on the quark coupling $g'_q$ as a function of $m_{W'}$ for $g'_\tau = 0.35\,,0.65\,,1$ at the 13 TeV LHC and $\mathcal{L}=30\,, 300\,, 3000\, \text{fb}^{-1}$ (solid black curves) superimposed with the 1$\sigma$ bands explaining the $R(D)$ (blue) and $R(D^\ast)$ (green) anomalies. Also shown are the recast inclusive limits from ATLAS at 13 TeV (dotted red curve), CMS at 13 TeV (dashed purple curve) and CMS at 8 TeV (dot-dashed magenta curve).}
\label{fig:limits}
\end{center}
\end{figure}

In Fig. \ref{fig:limits} we show the projected sensitivity of the 13 TeV LHC using our analysis for 30, 300, and 3000 $\text{fb}^{-1}$ of luminosity in the $g'_q-m_{W'}$ plane overlaying the 1$\sigma$ fits to the $R(D)$ and $R(D^\ast)$ measurements for representative choices of $g'_\tau$. Points above the black lines have a projected significance equal to or higher than 3$\sigma$. A large portion of the best fit bands should be within reach with current data and most of the remaining portion should likewise be by the end of Run 4. 

\begin{widetext}

\begin{table}[htbp]
\caption {Selection efficiencies in \% and the total number of events per 100 fb$^{-1}$ for the SM background processes. See the text and Table \ref{tab:selections} for details.}
\label{tab:cutefficiencySMnew}

\begin{tabular}{clllllll}\hline
{} &       $p_T(\tau)$ &             $p_T(b)$ & $e/\mu$ veto & $E_T^{\text{miss}}$ &      $|\Delta \phi (\tau_h,E_T^{\text{miss}})|$ & $N/(\text{100 fb}^{-1})$ \\\hline
$t\bar{t}$    & $3.29 \pm 0.0056$  &   $49.8 \pm 0.087$ &   $71.8 \pm 0.11$ &  $12.4 \pm 0.096$ &  $24.8 \pm 0.36$& $7.41\times 10^{3}$ \\
mono-$t$   & $1.13 \pm 0.0035$ &   $40.4 \pm 0.0035$ &   $90.5 \pm 0.14$ &   $2.55 \pm 0.082$ &   $31.4 \pm 1.5$&$5.95\times 10^{2}$ \\
$W+j$    &   $2.97 \pm 0.0094$ &   $8.4 \pm 0.09$ &   $94.1 \pm 0.26$ &    $9.65 \pm 0.34$ &     $23 \pm 1.6$ &$2.61\times 10^{3}$\\
DY$+j$      &    $2.87 \pm 0.014$ &    $14.1 \pm 0.18$ &   $97.4 \pm 0.21$ &  $6.45 \pm 0.34$ &   $32.4 \pm 2.5$ &$1.37\times 10^{3}$\\
$WW$    & $0.575 \pm 0.0042$ &  $7.66 \pm 0.19$ &    $92.5 \pm 0.7$ &  $6.63 \pm 0.68$ &   $21.6 \pm 4.4$&$3.80\times 10^{1}$ \\
$WZ$     &    $0.638 \pm 0.0071$&  $11.9 \pm 0.36$ &   $93.2 \pm 0.82$ & $6.7 \pm 0.84$ &     $39 \pm 6.3$ &$1.08\times 10^{2}$\\
$ZZ$      &        $0.673 \pm 0.012$&   $17.8 \pm 0.67$ &    $92.7 \pm 1.1$ &     $6 \pm 1$ &   $65.6 \pm 8.4$&$1.77\times 10^{1}$ \\
\hline
&&&&&Total Background&$1.22\times 10^{4}$\\
\end{tabular}
\end{table}

\begin{table}[htbp]
\caption {Selection efficiencies in \%, and the total number of events and significance per 100 fb$^{-1}$ for the signal for a range of $M_{W'}$ masses and $g'_q=g'_\tau=0.1$. See the text and Table \ref{tab:selections} for details.}
\label{tab:cutefficiencynw}

\begin{tabular}{cllllllll}\hline
{} &               $p_T(\tau)$ &             $p_T(b)$ & $e/\mu$ veto & $E_T^{\text{miss}}$ &      $|\Delta \phi (\tau_h,E_T^{\text{miss}})|$& $N/(\text{100 fb}^{-1})$& $\frac{S}{\sqrt{S+B}}$ \\\hline
200 GeV  &   $8.34 \pm 0.081$ &    $18.1 \pm 0.39$ &   $99.6 \pm 0.15$ &   $5.97 \pm 0.57$ &   $25.7 \pm 4.3$&$1.80\times 10^{2}$& 1.62 \\
250 GeV  &      $13.9 \pm 0.15$ &     $17.2 \pm 0.44$ &  $99.9 \pm 0.078$ &   $15.1 \pm 1$ &     $49 \pm 3.6$ &$5.98\times 10^{2}$&5.30\\
300 GeV  &    $18.3 \pm 0.24$ &     $17.4 \pm 0.56$ &   $99.9 \pm 0.13$ &   $28.6 \pm 1.6$ &   $69.2 \pm 3.1$& $1.06\times 10^{3}$&9.26\\
350 GeV  &    $22.2 \pm 0.27$ &     $17.1 \pm 0.52$ &   $99.7 \pm 0.19$ &   $37.6 \pm 1.6$ &   $67.8 \pm 2.5$&$8.88\times 10^{2}$&7.77 \\
500 GeV  &    $27.5 \pm 0.31$ &     $19.6 \pm 0.52$ &  $99.9 \pm 0.089$ &   $61.5 \pm 1.4$ &   $77.5 \pm 1.6$&$5.63\times 10^{2}$&5.00 \\
750 GeV  &    $31.5 \pm 0.34$ &     $21.7 \pm 0.54$ &   $99.7 \pm 0.16$ &   $79.1 \pm 1.2$ &   $82.4 \pm 1.2$&$1.55\times 10^{2}$&1.40 \\
1000 GeV &   $32.8 \pm 0.37$ &     $21.6 \pm 0.56$ &   $99.7 \pm 0.17$ &   $87.2 \pm 0.98$ &   $83.4 \pm 1.2$&$4.33\times 10^{1}$&0.39 \\\hline
\end{tabular}
\end{table}
\end{widetext}

Once $g'_\tau$ drops below about 0.35 we begin loosing the sensitivity of current data due to the small branching ratio to leptons, and the projected bounds completely disappear for the maximum luminosity below about $g'_\tau=0.03$. Increasing $g'_\tau$ would, vice versa, improve the reach until the branching ratio to leptons is saturated to be almost 1, beyond which the limits are unaffected but the $B$-anomaly bands move down to lower values of $g'_q$ to compensate. We remind the reader that the limits we impose are on the couplings assumed in Eq. \eqref{eq:lag} which, depending on the full theory, might have an upper bound comparable to our limits and should be read with this in mind.

Also shown are the inclusive limits (without the $b$-tag) recast from the ATLAS search using 13 TeV of energy and 36.1 $\text{fb}^{-1}$ of data \cite{Aaboud:2018vgh}, the CMS search using 13 TeV and 35.9 $\text{fb}^{-1}$ \cite{CMS:2018vff}, and the CMS search at 8 TeV with 19.7 $\text{fb}^{-1}$ \cite{Khachatryan:2015pua}. The 13 TeV CMS search provides stronger limits than our exclusive analysis for masses larger than about 500 GeV and weaker limits below that. The ATLAS 13 TeV limits, on the other hand, only starts competing with the exclusive limits at about 700 GeV. Note that the ATLAS limits appear to be significantly weaker than the CMS limits at the same energy and provide similar reach to the CMS limits at 8 TeV.



\section{Conclusion}
\label{sec:conclusion}

Recently reported $B$ anomalies by Babar, Belle and LHCb can be explained by  invoking new  $W^\prime$  gauge bosons  which couples preferentially to some of the second and third generation fermions. We have studied the LHC current and future reach for a simplified $W^\prime$ model that couples only to bottom quarks, charm quarks, and $\tau$-flavor leptons which is reponsibe for an enhanced charged current $b\rightarrow c\tau\nu$ that explains the $R(D)$ and $R(D^\ast)$ excesses.  We found that such a  $W^\prime$, which does not couple to the first generation fermions, can be produced from  $gc$, $gg$ and $cb$ fusion processes. 

This novel  production of  $W^\prime$  can lead to a $b$-jet along with  $\tau+\nu$ in the final state  which allows us to  use (i)  the existing inclusive analysis and (ii)  a new exclusive  $b+\tau_h+E^{\text{miss}}_T$ final state reported here to search for $W^\prime$ at the LHC. Our detailed study of the background and the signal showed that the presence of the $b$-jet in the exclusive final state is quite effective in reducing the SM background.  Further, the appearance of a $b$-jet in the final state would establish the coupling of the $W^\prime$ with the third generation fermions which is crucial for an explanation of the anomaly. 

The inclusive analyses from CMS and ATLAS have already constrained the $W^\prime$ mass down to 300 GeV and 500 GeV respectively. However, utilizing the cuts developed in our study we showed  that  the reach can be improved for the mass range $200-500$ GeV. We  showed that the mass range 250 - 750 GeV can be probed at $\sim 5\sigma$ level  even with 100 fb$^{-1}$ luminosity for $g_b=0.1$ and $g_\tau=0.1$. Consequently, a large region of parameter space which explains the anomaly can be investigated at the LHC directly due to the presence of the $b$-jet in the final state. The results of our analysis can be applied to a $W^\prime$ model with  additional decay channels after applying proper scaling arising from the $W^\prime$ branching ratios. We also noticed that the $b+\tau+\nu$ final state performs better for a $W^\prime$ search  compared to the recently reported $\geq 2\, {\rm jet} (\geq 1\, b)+ 2 \mu$ final state  for  a  $Z^\prime$ search (coupled to the second third generation fermions)  due to the existence of missing energy  in the former case~\cite{Dalchenko:2017shg}.




\section*{Acknowledgements}
We would like to thank Richard Ruiz for valuable advice and guidance on NLO calculations. B.D. and M. A. are supported in part by the DOE grant DE-SC0010813. We thank the constant and enduring financial support received for this project from the faculty of science at Universidad de los Andes (Bogot\'a, Colombia), the administrative department of science, technology and innovation of Colombia (COLCIENCIAS) through the Grant 111577657253, and  Sostenibilidad-UdeA.
\bibliographystyle{utphys}
\bibliography{Wp}
\end{document}